\documentclass[aps,prb,showpacs,twocolumn]{revtex4}
\newcommand\llangle{\langle\!\langle}
\newcommand\rrangle{\rangle\!\rangle}
\def\today{14 October 2012}

\usepackage{graphicx}
\begin{document}

\title
{

\begin{minipage}[t]{7.0in}
\scriptsize
\begin{quote}
\leftline{{\it Phys. Rev. B}, in press.
}
\raggedleft {\rm arXiv:1204.6499}
\end{quote}
\end{minipage}
\medskip

Collective Excitations and Stability of the Excitonic Phase in the Extended 
Falicov--Kimball Model}

\author{D. I. Golosov}
\email{golosov@mail.biu.ac.il}
\affiliation{Department of Physics and the Resnick Institute, Bar-Ilan 
University, Ramat-Gan 52900, Israel.}

\date{\today}

\begin{abstract}
We consider the excitonic insulator state (often associated with 
electronic ferroelectricity), which
arises on the phase diagram of an extended spinless 
Falicov--Kimball model (FKM) at half-filling.
Within the Hartree--Fock approach, we calculate the spectrum of  low-energy 
collective excitations in this state up to second order in the narrow-band 
hopping and/or hybridisation. This allows to probe the mean-field stability of
the excitonic insulator. The latter is found to be unstable when the
case of the pure FKM (no hybridisation with
a fully localised band) is approached. The instability is
due to the presence of another, lower-lying ground state and {\it not} to
the degeneracy of the excitonic phase in the pure FKM. The excitonic phase,
however, may be stabilised further away from the pure FKM limit.
In this case, the low-energy excitation spectrum  contains new information 
about  the properties of the excitonic condensate (including the 
strongly suppressed critical temperature).

\typeout{polish abstract}
\end{abstract}
\pacs{71.10.Fd,  71.28.+d, 71.35.-y,  71.10.Hf}
\maketitle

\section{INTRODUCTION}

The Falicov--Kimball model (FKM)\cite{FK} was introduced more than forty years
ago with an objective to describe metal-insulator transitions and 
mixed-valence phenomena in certain hexaborides and oxides. It attracted 
an impressive
amount of work within both solid state theory\cite{Khomrev,Zlatic} 
and mathematical physics\cite{Gruber},
dealing with all aspects of the ground state properties and the phase diagram.
Early on, various extensions of the original  model were 
proposed\cite{Khomrev,Kaplan} in
order to more adequately describe the physics and/or phenomenology of
specific systems. In recent years, much attention has been paid to the
possibility of {\it electronic ferroelectricity} in an appropriately
extended Falicov--Kimball model (EFKM)\cite{Portengen}. This phenomenon 
is intimately
related to the notion of excitonic insulator, which was introduced earlier
in a somewhat different context\cite{Keldysh,Jerome,Kohn,Wachter90} (see also
Refs. \onlinecite{KaplanMahanti,Kikoin83,Kikoin00}). Finding an actual
electronic ferroelectric might be important technologically, and the
experimental search is ongoing\cite{expferro}. 
In addition, it has been suggested that the EFKM might
be relevant for some  systems of current interest, including 
manganates\cite{Rama} and ${\rm URu_2 Si_2}$\cite{Dubi}.

Various approximate methods\cite{Czycholl99,Farkasovsky08,Czycholl08,Zenker10} 
(as well as numerical techniques\cite{Batista02,Farkasovsky02}) were used
to determine the phase diagrams of both pure and extended FKM and the carrier 
spectral 
properties in the ground state. Yet it appears that the present article 
reports the first
systematic study of the low-energy collective excitations in the excitonic
insulator (or excitonic condensate) state of the EFKM. Even
though restricted by the Hartree--Fock mean field approach, the results
shed light on the nature of much discussed instability of the excitonic 
state in the
limit of pure FKM. In the case of EFKM, we obtain the excitation spectra in
the excitonic state, and thereby
identify the underlying energy scales. Importantly, this also allows to
probe the mean-field stability of the uniform excitonic insulator. Until now,
this latter issue has been addressed only variationally, by considering
a finite set of competing mean field ground states.

The spinless Falicov--Kimball model proper involves fermions $d_i$ and
$c_i$ in the localised and itinerant bands, interacting via a Coulomb
repulsion $U$ on-site:
\begin{equation}
{\cal H}=-\frac{t}{2}\sum_{\langle i j \rangle} \left(c^\dagger_i c_j +
c^\dagger_j c_i \right) + E_d \sum_i d^\dagger_i d_i +U \sum_i c^\dagger_i
d^\dagger_i d_i c_i\,,
\label{eq:FKM}
\end{equation}
where $E_d$ is the bare energy of the localised band. We choose the units
where the hopping amplitude $t$ and the period $a$ of the ($d$-dimensional
hypercubic) lattice are equal to unity. 
Presently, we consider the half-filled ($n=1$) case at zero temperature. 

Historically, the {\it raison d'\^{e}tre} of the Falicov--Kimball model is
the presence of an intermediate valence regime, whereby the two original bands
are partially filled.
It is well-known\cite{Khomskii76} that within a broad
range of parameter values in this regime,  the lowest-energy 
spatially uniform mean-field solution  
is characterised by an interaction-induced spontaneous hybridisation, 
$\Delta = \langle c^\dagger_i d_i \rangle$. Hartree-Fock decoupling of the 
equations of motion for the Green's functions $\llangle c_{\vec{k}}
c^\dagger_{\vec{k}}\rrangle$ and $\llangle d_{\vec{k}}
c^\dagger_{\vec{k}}\rrangle$ yields 
\begin{equation}
\Delta = \frac{1}{N} \sum_{\vec{k}} \Delta_{\vec{k}}\,,\,\,\,
\Delta_{\vec{k}}\equiv \langle c^\dagger_{\vec{k}}d_{\vec{k}}\rangle=\frac{U \Delta}{\sqrt{\xi_{{\vec{k}}}^2+4 U^2 |\Delta|^2}}
\label{eq:Delta}
\end{equation}
for the hybridisation and
\begin{equation}
n_d=\frac{1}{N}\sum_{\vec{k}} n^d_{\vec{k}}\,,\,\,\,
n^d_{\vec{k}} \equiv \langle d^\dagger_{\vec{k}}d_{\vec{k}}\rangle= 
\frac{1}{2}-\frac{\xi_{{\vec{k}}}}{2\sqrt{\xi_{{\vec{k}}}^2+4 U^2 |\Delta|^2}}
\label{eq:nd}
\end{equation}
for the localised-band occupancy $n_d$. Here, $N$ is the number of the lattice 
sites and $\xi_{\vec{k}}=E_{rd} -\epsilon_{\vec{k}}$ with the renormalised
 relative energy of the localised band, $E_{rd}=E_d + U (1- 2 n_d)$, and the 
tight-binding dispersion, $\epsilon_{\vec{k}}=
-\sum_{\alpha=1}^d \cos k_\alpha$.

This mean-field solution has its filled and empty quasiparticle bands separated
by a hybridisation gap of $2U |\Delta|$, corresponding to an
excitonic insulator. If the two original bands have opposite parity and 
${\rm Re} \Delta \neq 0$, it has a spontaneous dipole moment (electronic 
ferroelectricity)\cite{Portengen,Batista02}. However, the uniform excitonic 
insulator state in the pure
FKM is unstable\cite{Subrahmanyam,Farkasovsky02}, which has largely been 
attributed\cite{Zlatic,Subrahmanyam} 
to fluctuations due to the local continuous 
degeneracy. The latter, in turn, is associated with the phase of each of 
the operators $d_i$ in Eq. 
(\ref{eq:FKM}), and hence with the phase of $\langle c^\dagger_i d_i \rangle$
on-site.

This strong degeneracy can be broken, and the stability of the  
excitonic insulator may eventually be restored, by any of the terms in the 
perturbation,
\begin{eqnarray}
\delta {\cal H}=&&-\frac{t^\prime}{2}\sum_{\langle i j \rangle} d^\dagger_i d_j + V_0 \sum_i c^\dagger_i d_i -\frac{V_1}{2}\sum_{\langle i j \rangle}\left( 
c^\dagger _i d_j + c^\dagger_j d_i \right)-\nonumber \\
&& -\frac{V_2}{2} 
\sum_{\langle i j \rangle}\left\{ (\vec{R}_j-\vec{R_i})\cdot \vec{\Xi}
\right\}
\left(c^\dagger _i d_j - c^\dagger_j d_i \right) + {\rm H. c.}\,,
\label{eq:pert}
\end{eqnarray}
where $t^\prime$ is the $d$-band hopping and $V_0$, bare  on-site
hybridisation.
$V_1$ ($V_2$) is the spatially-even (odd) nearest-neighbour hybridisation,
as appropriate for the case where the two original bands have the same 
(opposite) parity. $\vec{R}_i$ is the radius-vector of a site $i$, and 
$a\vec{\Xi}=\sum_{\alpha=1}^{d}\hat{\mathbf x}_\alpha$, sum of 
Cartesian unit vectors.

When the Falicov--Kimball model is {\it extended} by Eq.(\ref{eq:pert}), 
it ceases to
be exactly soluble. Identifying its ground states and constructing the 
phase diagram constitute the subject of a broad on-going research 
effort\cite{ZenkerBatista,current}. 
In the present article, we address the issue of stability of the excitonic 
insulating state with a uniform real  $\Delta$, and its low-energy 
properties at the mean-field level (ignoring higher-order correlations). 

\section{LOW-ENERGY COLLECTIVE EXCITATIONS IN THE EXCITONIC PHASE}

We consider the particle-hole 
excitations of the excitonic insulator state, which in general are given by
\begin{eqnarray}
{\cal X}_{\vec{q}}=&&\frac{1}{\sqrt{N}} \sum_{\vec{k}}\left\{
F_+(\vec{k},\vec{q})\, c^\dagger_{\vec{k}}d_{\vec{k}+\vec{q}}+
F_-(\vec{k},\vec{q})\, d^\dagger_{\vec{k}}c_{\vec{k}+\vec{q}}+\right. 
\nonumber \\
&&\left. +F_c(\vec{k},\vec{q})\, c^\dagger_{\vec{k}}c_{\vec{k}+\vec{q}}+
F_d (\vec{k},\vec{q})\,d^\dagger_{\vec{k}}d_{\vec{k}+\vec{q}} \right\}\,,
\label{eq:wave}
\end{eqnarray}
where the four quantities $F_i$ are arbitrary functions
of $\vec{k}$ and $\vec{q}$. The energies $\omega$ and the wave functions of
collective modes satisfy the secular equation,
\begin{eqnarray}
\left[{\cal X}_{\vec{q}}, {\cal H} + \delta{\cal H} \right]
&\stackrel{\rm eff}{=}&\omega_{\vec{q}} {\cal X}_{\vec{q}}\,,
\label{eq:secular} 
\\
&\,\,& \nonumber
\end{eqnarray}
where the label ``eff'' implies Hartree--Fock decoupling on the l. h. s.. 
Upon substituting Eq. (\ref{eq:wave}) into (\ref{eq:secular}), 
we find a system of four equations for the functions $F_i$. For 
example,
collecting the terms with $c^\dagger_{\vec{k}}d_{\vec{k}+\vec{q}}$ and those with $d^\dagger_{\vec{k}}c_{\vec{k}+\vec{q}}$ yields respectively
\begin{widetext}
\begin{eqnarray}
(\omega - {\xi}_{\vec{k}}- t^\prime 
\epsilon_{\vec{k}+\vec{q}}) F_+(\vec{k},\vec{q})+ (U {\Delta}^*-
V_{\vec{k}+\vec{q}})F_c(\vec{k},\vec{q})
-(U {\Delta}^*-V_{\vec{k}})
F_d(\vec{k},\vec{q})&=&U\left[A_a(\vec{q})+A_b(\vec{q})\right]\,,
\label{eq:F+}\\
(\omega + {\xi}_{\vec{k}+\vec{q}}+ t^\prime 
\epsilon_{\vec{k}}) F_-(\vec{k},\vec{q})- (U {\Delta}-
V_{\vec{k}}^*)F_c(\vec{k},\vec{q})
+(U {\Delta}-V_{\vec{k}+\vec{q}}^*)
F_d(\vec{k},\vec{q})&=&U\left[-A_a(\vec{q})+A_b(\vec{q})\right]\,.\label{eq:F-}
\end{eqnarray}
Similarly,
\begin{eqnarray}
(U {\Delta}-V^*_{\vec{k}+\vec{q}}) F_+(\vec{k},\vec{q})-
(U {\Delta}^*-V_{\vec{k}}) F_-(\vec{k},\vec{q})+
(\omega + {\xi}_{\vec{k}+\vec{q}}-\xi_{\vec{k}})F_c(\vec{k},\vec{q})
&=&UA_c(\vec{q})\,,
\label{eq:Fc}\\
-(U {\Delta}-V^*_{\vec{k}}) F_+(\vec{k},\vec{q})
+(U {\Delta}^*-V_{\vec{k}+\vec{q}}) F_-(\vec{k},\vec{q})+
(\omega- t^\prime \epsilon_{\vec{k}+\vec{q}}+t^\prime 
\epsilon_{\vec{k}})F_d(\vec{k},\vec{q})&=&UA_d(\vec{q})\,.
\label{eq:Fd}
\end{eqnarray}
Here, the quantities $A_i$ originate from the interaction term in Eq.(\ref{eq:FKM}) and obey the self-consistency conditions,
\begin{eqnarray}
A_a(\vec{q})&=&\frac{1}{2N} \sum_{\vec{p}}\left\{\left[F_+(\vec{p},\vec{q})+F_-(\vec{p},\vec{q})\right] (\tilde{n}^d_{\vec{p}}+
\tilde{n}^d_{\vec{p}+\vec{q}}-1)+F_c(\vec{p},\vec{q})(\tilde{\Delta}^*_{\vec{p}+\vec{q}}+\tilde{\Delta}_{\vec{p}})-F_d(\vec{p},\vec{q})(\tilde{\Delta}_{\vec{p}+\vec{q}}+\tilde{\Delta}^*_{\vec{p}})\right\}\,,
\label{eq:Aa}\\
A_b(\vec{q})&=&\frac{1}{2N} \sum_{\vec{p}}\left\{\left[F_+(\vec{p},\vec{q})-F_-(\vec{p},\vec{q})\right] (\tilde{n}^d_{\vec{p}}+
\tilde{n}^d_{\vec{p}+\vec{q}}-1)+F_c(\vec{p},\vec{q})(\tilde{\Delta}^*_{\vec{p}+\vec{q}}-\tilde{\Delta}_{\vec{p}})+F_d(\vec{p},\vec{q})(\tilde{\Delta}_{\vec{p}+\vec{q}}-\tilde{\Delta}^*_{\vec{p}})\right\}\,,
\label{eq:Ab}\\
A_c(\vec{q})&=&\frac{1}{N} \sum_{\vec{p}}\left\{F_+(\vec{p},\vec{q}) \tilde{
\Delta}_{\vec{p}}-F_-(\vec{p},\vec{q}) \tilde{
\Delta}^*_{\vec{p}+\vec{q}}+F_d(\vec{p},\vec{q})\left[\tilde{n}^d_{\vec{p}}-
\tilde{n}^d_{\vec{p}+\vec{q}}\right]\right\}\,,
\label{eq:Ac}\\
A_d(\vec{q})&=&\frac{1}{N} \sum_{\vec{p}}\left\{-F_+(\vec{p},\vec{q}) \tilde{
\Delta}_{\vec{p}+\vec{q}}+F_-(\vec{p},\vec{q}) \tilde{
\Delta}^*_{\vec{p}}-F_c(\vec{p},\vec{q})\left[\tilde{n}^d_{\vec{p}}-
\tilde{n}^d_{\vec{p}+\vec{q}}\right]\right\}\,,
\label{eq:Ad}
\end{eqnarray}
\end{widetext}
$V_{\vec{k}}$ is the Fourier component of the bare hybridisation,
\begin{equation}
V_{\vec{k}}=\left\{\begin{array}{ll}
V_0+V_1 \epsilon_{\vec{k}},\,\,\,&\mbox{even}\\ 
{\rm i} V_2 \lambda_{\vec{k}},\,&\mbox{odd}\end{array}\,,
\right.
\,\,\,\lambda_{\vec{k}}=-\sum_{\alpha=1}^d \sin k_\alpha.
\end{equation}
(depending on the relative parity of the orbitals).
The tilde accents in Eqs. (\ref{eq:Aa}--\ref{eq:Ad}) reflect the fact
that in the 
presence of $\delta{\cal H}$, the r.h.s. of Eqs. 
(\ref{eq:Delta}--\ref{eq:nd}) are trivially 
modified, and the corrected 
expressions [see Appendix, Eqs. 
(\ref{eq:Deltat}--\ref{eq:ndt})] should be 
used here.
Ideally, one should now solve Eqs. (\ref{eq:F+}--\ref{eq:Fd}) 
and 
substitute 
the explicit expressions for $F_i(\vec{q})$ into Eqs. 
(\ref{eq:Aa}-\ref{eq:Ad}). 
Zeroes of the determinant ${\cal D}(\omega,\vec{q})$ of the resultant 
system of four linear homogeneous 
equations for $A_i(\vec{q})$ would then yield the spectrum of collective 
excitations for a given
momentum $\vec{q}$ (cf. Ref. \onlinecite{Jerome}). Since we are interested in the 
low-energy excitations only,
it is possible to follow a simpler route as sketched below.

For the pure FKM with $\delta {\cal H}=0$ and real $\Delta$, these homogeneous
equations  for $A_i(\vec{q})$
have an $\omega_{\vec{q}} \equiv 0$
solution with $A_a=-\Delta$ and $A_b=A_c=A_d=0$, corresponding to\cite{longitudinal}
\begin{equation}
{\cal X}^{(0)}_{\vec{q}}=\frac{1}{\sqrt{N}}\sum_{\vec{k}}d^\dagger_{\vec{k}}d_{\vec{k}+\vec{q}}
\label{eq:longitudinal}
\end{equation}
[which the reader can verify by substituting into Eq. (\ref{eq:secular})].
The presence of an entire branch of zero-energy excitations is an expected
consequence of the local continuous degeneracy as discussed in the Introduction
above.
For the EFKM with sufficiently weak perturbation, Eq. (\ref{eq:pert}), this 
branch acquires a small but finite energy that can be found by expanding
the equations for $F_i$ and $A_i$ in powers of $\omega$, $V_i$, and $t^\prime$.

\section{LEADING-ORDER INSTABILITY OF THE EXCITONIC INSULATOR}

Assuming that $\Delta$  (but not necessarily
$\tilde{\Delta}_{\vec{k}}$, see Ref. \onlinecite{Portengen}) is real, we find to 
leading order:
\begin{equation}
{\cal D}(\omega,\vec{q})=D_\omega(\vec{q})\cdot \left( \frac{\omega}{U\Delta} 
\right)^2 + M_{11}(\vec{q}) \cdot D_0 (\vec{q}).
\label{eq:D}
\end{equation}
Here, the first term comes from expanding ${\cal D}(\omega,\vec{q})$ of the 
pure FKM in powers
of $\omega$:
\begin{eqnarray}
D_\omega(\vec{q})&=&-[Y_1(\vec{q})]^2-Z_2(\vec{q})Y_1(\vec{q})-\left[2Y_0(\vec{q})-
Z_1(\vec{q})\right]^2\,,\nonumber \\
Y_n(\vec{q})&=&\frac{U^{1-n}}{\Delta^n N}\sum_{\vec{k}}
\frac{\xi_{\vec{k}}^n {\rm Re} \Delta_{\vec{k}}}{\xi_{\vec{k}+\vec{q}}-\xi_{\vec{q}}}\,.
\label{eq:Yn}
\end{eqnarray}
Since the two quantities 
\begin{eqnarray}
\!\!Z_1(\vec{q})&=&Y_2(\vec{q})+\frac{1}{2\Delta}(1-2n_d)+4Y_0(\vec{q})\,,
\nonumber \\
\!\!Z_2(\vec{q})&=&-\frac{E_{rd}}{U\Delta^2}-Y_3(\vec{q})-2-4Y_1(\vec{q})+
\frac{2}{U\Delta^2N}\sum_{\vec{k}}\xi_{\vec{k}}n^d_{\vec{k}}
\nonumber 
\end{eqnarray}
vanish in the long-wavelength limit, $q \rightarrow 0$, we find $D_\omega(\vec{q}=0)<0$. Likewise, one can show that $D_\omega(\vec{q})$ at $\vec{q}=\vec{Q}_0\equiv
\{\pi,\pi(,\pi)\}$ is given by\cite{sign}
\begin{eqnarray}
 D_\omega(\vec{Q}_0)=&-&\left[\frac{E_{rd}Y_0(\vec{Q}_0)}{U\Delta}-\frac{1}{2}\right]^2
-4[Y_0(\vec{Q}_0)]^2- \nonumber \\
&-&\frac{1}{\Delta^2N^2}
\sum_{\vec{k}}\epsilon_{\vec{k}}n^d_{\vec{k}}\sum_{\vec{p}}\frac{n^d_{\vec{p}}}
{\epsilon_{\vec{p}}}<0\,,
\nonumber
\end{eqnarray}
and numerically we find that $D_\omega$ remains negative throughout 
the Brillouin
zone (BZ). 

To leading order in perturbation, the determinant ${\cal D}$ for the EFKM 
[see Eq. (\ref{eq:D})] at $\omega=0$ factorises into a product of the $3\times 3$ diagonal minor 
(corresponding to $A_b$, $A_c$, and $A_d$),
\begin{eqnarray}
&&D_0(\vec{q})=\left[4-2Z_2(\vec{q})\right]\left\{[Y_1(\vec{q})]^2+
4[Y_0(\vec{q})]^2\right\}+ \nonumber \\
&&+2\left\{1-[Z_1(\vec{q})]^2\right\}Y_1(\vec{q})
-8Z_1(\vec{q})\,Y_0(\vec{q})\,,
\label{eq:D0}
\end{eqnarray}
and the diagonal matrix element corresponding to $A_a$,
\[ 
\!\!M_{11}\!=\! \frac{t^\prime(d+ \epsilon_{\vec{q}}) }{U\Delta^2Nd}
\sum_{\vec{k}}\! \epsilon_{\vec{k}}
n^d_{\vec{k}} +\left\{ \begin{array}{ll}\!\!
 {\displaystyle \frac{V_0\!+V_1 E_d} {U\Delta},} &\mbox{even,} \\\!\!
~ & ~ \\ \!\!
{\displaystyle \frac{V^2_2}{N}
\!\sum_{\vec{k}}\!\frac{(\lambda_{\vec{k}}+
\lambda_{\vec{k}+\vec{q}})^2n^d_{\vec{k}}}{U\Delta^2(\xi_{\vec{k}}
-\xi_{\vec{k}+\vec{q}})},}&\mbox{odd.} \end{array} \right.
\] 
When $V_i=0$, the latter quantity (and hence\cite{Jerome,recover} $\omega_{\vec{q}}$) vanishes at $\vec{q}=0$,
corresponding to the $U(1)$ remaining degeneracy for $t^\prime \neq 0$ 
that was discussed 
earlier\cite{Batista04}.
When $t^\prime=V_{0,1}=0$, a different Goldstone mode is present at 
$\vec{q}=\vec{Q}_0$. It is due to the degeneracy of the excitonic insulator 
with respect to assigning arbitrary opposite phases, $\pm \varphi$, to 
$\langle c^\dagger_i d_i \rangle$ in a checker-board order while keeping
$n_d$ and $|\langle c^\dagger_i d_i \rangle|=\Delta$ unchanged\cite{checker}. 
This 
can be verified directly by analysing the mean field
equations at $V_2 \neq 0$ for the 2-sublattice case. In all other cases, 
the appropriate choice of signs ($t^\prime,V_{0}\Delta,V_1 E_d \Delta \leq 0$) 
ensures that 
$M_{11}$ is negative throughout the BZ.   

At $\vec{q}=0$, the quantity $D_0$ is proportional to the derivative, $\partial \Delta/\partial V_0$, calculated in the unperturbed case, and is therefore
negative (see Appendix \ref{app:signD0}). The equation ${\cal D}(\omega, \vec{q})=0$ then yields a positive 
$\omega^2$ and hence a stable spectrum in the
long-wavelength limit\cite{recover}. It is, however, easy to show that  at $\vec{q}=\vec{Q}_0$,
\begin{eqnarray}
D_0(\vec{Q}_0)=&&\left\{\left[\frac{2 E_{rd}}{U\Delta}Y_0(\vec{Q}_0) -1\right]^2+
16[Y_0(\vec{Q}_0)]^2\right\} \times \nonumber \\
&& \times \frac{1}{2U\Delta^2 N}\sum_{\vec{k}}\epsilon_{\vec{k}}n^d_{\vec{k}}
\end{eqnarray}
and is {\it positive}\cite{sign,zero}. This gives imaginary $\omega$ at $\vec{q}=\vec{Q}_0$, implying that the 
excitonic insulator is always unstable when the pure FKM limit is 
approached.

Contrary to earlier suggestions \cite{Zlatic,Subrahmanyam}, this type of 
instability is due to the presence of a lower lying ground state, and
{\it not} to the local degeneracy of the excitonic phase in the FKM. This
agrees with those variational results\cite{Czycholl99,Farkasovsky08} 
that suggest that a lower-energy ordered state is favoured in the pure
FKM. The excitonic insulator is then stabilised by a small but finite 
perturbation via a second-order phase
transition\cite{Czycholl99,Farkasovsky08}, whereas a more exotic 
behaviour\cite{exotic} would have been expected should the vanishing 
$\langle c^\dagger_i d_i \rangle$ in the pure 
FKM case\cite{Farkasovsky02} be due to the local degeneracy. The latter,
on the contrary, {\it masks} the instability in the unperturbed case (by
causing the spectrum to vanish identically). Hence in order to see this
instability, one has to allow for an infinitesimal perturbation, as 
outlined above. 

\section{STABILISATION OF THE EXCITONIC PHASE BEYOND THE LEADING ORDER}
 
When effects of subleading order in perturbation are taken into account, the 
structure of  Eq. (\ref{eq:D}) is preserved. In the following, we will keep 
the most significant next-order (in $t^\prime$, $V_{0,1}$, and $V_2^2$)
corrections: (i) Corrections to the 
determinant  $D_0$, which is responsible for the leading-order instability. 
We will include these by substituting in Eq. (\ref{eq:D}) 
$D_0 \rightarrow \tilde{D}_0+D_1$, where $\tilde{D}_0$ is given by Eq. (\ref{eq:D0}) 
with corrected
expressions $\tilde{\Delta}_{\vec{k}}$ and $\tilde{n}^d_{\vec{k}}$ used
in the definitions of $Y_n$ and $Z_n$. Discussion of the quantity
$D_1(\vec{q})$ is relegated to Appendix \ref{app:D1}. 
(ii) In all terms in Eq. (\ref{eq:D}), we use corrected values of
$\Delta$ and $n^d$ [obtained when $\tilde{\Delta}_{\vec{k}}$ and 
$\tilde{n}^d_{\vec{k}}$ are used in Eqs. (\ref{eq:Delta}--\ref{eq:nd})]. 
This is
to account for the sometimes appreciable\cite{jacobian} change of $\Delta$ and $n^d$
caused by a weak perturbation.

\begin{figure}
\includegraphics{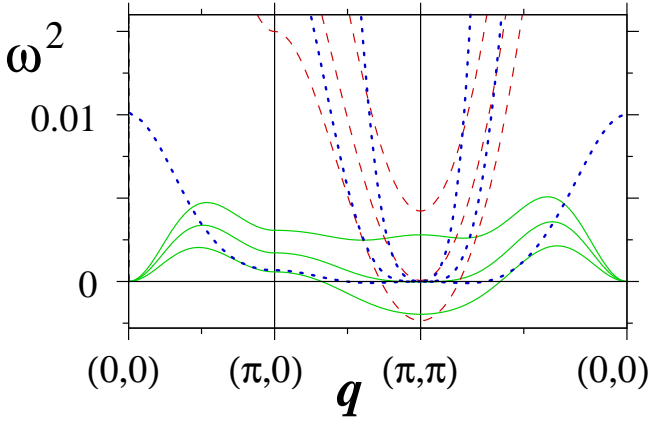}
\caption{\label{fig:spectra} (colour online) Effect of different 
perturbations on the
excitation spectrum of the 2D FKM, Eq.(\ref{eq:FKM}) with $U=2$, $E_d=0.4$. 
Solid, dashed, and dotted lines correspond, respectively, to the effects 
of $t^\prime$ (bottom to top:  $t^\prime=-0.04,-0.0565,-0.07 $), $V_0$ 
($V_0=-0.06,-0.074 ,-0.09 $), 
and $V_2$ ($V_2=0.115,0.249,0.35$). As the corresponding values of $\Delta$
vary between $0.214$ and $0.359$, the gap in the single-particle spectrum is
in all cases larger than $\omega_{\vec{q}}$ for any $\vec{q}$ in the BZ.}
\end{figure}

Numerical evaluation shows that the quantity $D_1$ at $\vec{q}=\vec{Q}_0$
typically is {\it negative}, implying that for perturbations larger than
a certain critical value $t^\prime_{cr}$ or $V_{i,cr}$ the spectrum may become 
real and the stability of the uniform excitonic phase with respect to the
excitations of the type of Eq. (\ref{eq:wave})
may be restored, as shown in Fig. \ref{fig:spectra} (see also Appendix
\ref{app:figdetail}). 
Applying our 
perturbative treatment to a finite $\delta {\cal H}$ 
is justifiable whenever the resultant critical values of $t^\prime$ or $V_i$ 
are small. The latter includes  requirements, $|V_{0,1}| \ll U \Delta$ and
$|V_i|, |t^\prime| \ll 1$. 

\begin{figure}
\includegraphics{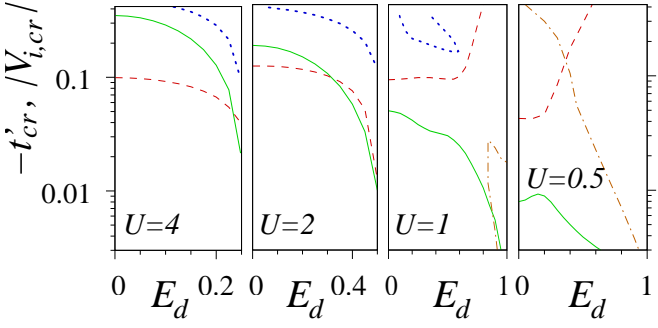}
\caption{\label{fig:critical} (colour online) Critical values of $t^\prime$ 
(solid line, see Appendix \ref{app:figdetail} for further details), 
$V_0$ (dashed), $V_1$ (dashed-dotted) and $V_2$ (dotted) in a 2D EFKM with 
$U= 0.5$ (weak
coupling regime, right panel), $U=1$ (intermediate coupling, second right), 
$U=2$ (second left), and $U=4$ (strong coupling, left panel). Results for $U=1$
suggest reentrant behaviour in $V_{1,2}$.}
\end{figure}

The critical values of $V_i$ and $t^\prime$ as 
functions of $E_d$ are plotted in Fig. \ref{fig:critical}. 
Wherever a direct comparison with earlier variational and numerical
results is possible (the $t^\prime \neq 0$ 
case\cite{Farkasovsky08,Batista04}), our approach yields an {\it increased} 
stability region for the excitonic
phase, with the disagreement becoming more pronounced as $|E_d|$ decreases.
This may be due
to the above extrapolation scheme being inexact, or  to a possible
importance of
more complex excitations [as opposed to Eq. (\ref{eq:wave})],
or to the presence of a first-order transition\cite{Czycholl08,Batista02} 
at small $E_d$. 

\section{DISCUSSION}

We shall now attempt to put our results into a broader prospective. 
Whenever the excitonic insulator behaviour is suggested, a distinction 
must be made as to whether the underlying physics
is that of FKM [with one of the bands being (nearly) flat] or of a generic
semimetal (semiconductor) with nested Fermi 
surfaces\cite{Keldysh,Jerome,Kohn,Wachter90}. Our results show that the difference is not
restricted to the parent bandstructure, but is also manifest in the 
observable properties of the system. Indeed, the presence of an entire branch
of low-energy excitations  is characteristic
only of the EFKM-based mechanism (cf. Ref. \onlinecite{Jerome}).

A physical difference between the two realisations of excitonic insulator
is indeed expected on very general grounds\cite{Kikoin83,Kikoinprivate}. 
In the case of a broad-band
semimetal (semiconductor), the excitonic (induced hybridisation) gap is
typically much smaller than the bandwidth of either band. The 
particle-hole pairing (exciton formation) is then essentially a Fermi-surface 
effect, involving only electrons and holes from the vicinities of the two Fermi
surfaces, therefore nesting is required\cite{Kohn}. In this case, the 
physics
of excitonic insulator is largely the same as that of a conventional
charge-density wave.  On the other hand, in the case of EFKM (sufficiently close to the
pure FKM limit), the width of the narrow band is smaller than the gap,
hence the entire narrow band is in principle accessible for excitonic 
pairing and the Fermi surface shape is no longer crucial. The situation is
then closer to the actual intermediate valence 
picture\cite{KaplanMahanti,Kikoin83} 
(especially where slower dynamics is concerned). We note that the
bandstructure of most compounds suggested  as 
candidates for excitonic insulating state\cite{Wachter90,KaplanMahanti,Kikoin83,Kikoin00,excitonicins}, includes 
a very narrow band
corresponding to a slightly perturbed FKM.

With increasing temperature, the long-range order of the excitonic insulator 
state in a semimetal is lost owing to the thermal excitations of electron-hole 
continuum. The critical temperature (which has the well-known snail-like 
profile as a function of band overlap\cite{Kohn} or of the interaction 
strength\cite{current} $U$)  is found from the gap equation\cite{Kohn}. 
It appears that the current view \cite{Czycholl08,Zenker10,current} is that a 
similar scenario 
should hold for the EFKM.

Our results suggest that the behaviour of an FKM-based excitonic insulator 
involves a second-order 
transition\cite{crossover} mediated by the low-energy excitations reported 
above\cite{firstorder}.
While even for the pure FKM case the gap equation, Eq. (\ref{eq:Delta}), has a
non-zero solution at low temperatures, $T < T_{\Delta} {\sim} U\Delta(T=0)
/ k_B$, the actual critical temperature $T_c$ for a weakly perturbed FKM 
does not simply follow the single-particle gap. Instead, the relevant energy 
scale is determined 
by the  collective excitations, {\it i. e.}, by the dominant 
term in Eq. (\ref{eq:pert}), resulting in $T_c \ll T_\Delta$. 
It is easy to see that $T_c$ 
will be roughly 
proportional to 
$\sqrt{|t^\prime|(|t^\prime|-|t^\prime_{cr}|)}$, 
$|V_2|\sqrt{V_2^2-V_{2,cr}^2}$, 
$\sqrt{|V_0|(|V_0|-V_{0,cr})}$, or $\sqrt{|V_1|(|V_1|-|V_{1,cr}|)}$. 

Finally, we summarise our conclusions that appear relevant in the context of 
experimental search for an excitonic insulator with a narrow (several meV) 
and a broad ($W\stackrel{>}{\sim} 0.1$ eV) bare carrier bands, and 
$U/W \stackrel{>}{\sim}0.2$. 
(i) $T_c$ in this 
case would not exceed several tens K, and could be much smaller.
(ii) A peculiar branch of excitations (``excitonic phonon''\cite{Kohn},
either acoustic or optical, largely undamped throughout the BZ), will be 
present 
in the infrared or even microwave range at $T<T_c$.
(iii) Since
these excitations involve mostly the 
narrow-band charge 
density waves,
Eq. (\ref{eq:longitudinal}), one can also expect to find static or dynamic 
charge and/or
orbital inhomogeneities in a region above $T_c$. 
While these predictions are based on a Hartree-Fock study of a spinless
model, qualitatively they are expected to persist in a more realistic
treatment.  


\acknowledgements
It is a pleasure to thank C. D. Batista, R. Berkovits,  
D. I. Khomskii, K. A. Kikoin, B. D. Laikhtman, and A. A. Sokolik 
for discussions.
This work was supported by the Israeli Absorption Ministry.


\appendix

%

\section {Corrected mean-field equations} 

In the presence of 
perturbation, Eq. (4), Eqs. (2-3) hold to leading order only. The exact
mean field equations read
\begin{equation}
\Delta = \frac{1}{N} \sum_{\vec{k}} \tilde{\Delta}_{\vec{k}}\,,\,\,\,
\tilde{\Delta}_{\vec{k}}=\frac{U \Delta-V_{\vec{k}}^*}{\sqrt{(\xi_{{\vec{k}}}
+t^\prime \epsilon_{\vec{k}})^2+4 |U\Delta-V_{\vec{k}}^*|^2}}
\label{eq:Deltat}
\end{equation}
for the hybridisation and
\begin{equation}
n_d=\frac{1}{N}\sum_{\vec{k}} \tilde{n}^d_{\vec{k}}\,,\,\,\,
\tilde{n}^d_{\vec{k}} = 
\frac{1}{2}-\frac{\xi_{{\vec{k}}}+t^\prime \epsilon_{\vec{k}}}{2
\sqrt{(\xi_{{\vec{k}}}
+t^\prime \epsilon_{\vec{k}})^2+4 |U\Delta-V_{\vec{k}}^*|^2}}
\label{eq:ndt}
\end{equation}
for the occupancy of the (almost) localised band.

\section{Sign of $D_0$ at $\vec{q}\rightarrow 0$}
\label{app:signD0}

Suppose that $\Delta$ ($ \neq 0$) and $n_d$ solve the mean field equations, 
Eqs. 
(\ref{eq:Delta}-\ref{eq:nd}), for the unperturbed FKM, Eq. (\ref{eq:FKM}).
Adding a weak on-site hybridisation $V_0$ [{\it i.e.},
a perturbation $\delta {\cal H}$, given by Eq. (\ref{eq:pert}) with
$t^\prime=V_{1,2}=0$] results in a small change of the 
mean-field parameters:
$\Delta \rightarrow \Delta+\delta \Delta$, $n_d \rightarrow n_d+\delta n_d$. 
Keeping the linear-order (in $\delta \Delta$, $\delta n_d$, and $V_0$) terms 
in Eqs. (\ref{eq:Deltat}-\ref{eq:ndt}) yields a system of two equations,
\begin{eqnarray}
\delta \Delta&=&\left(\delta \Delta -\frac{V_0}{U}\right) \frac{1}{N}
\sum_{\vec{k}}\frac{U \xi_{\vec{k}}^2}{(\xi_{\vec{k}}^2+4U^2 \Delta^2)^{3/2}}+
\nonumber \\
&&+2\delta n_d\frac{1}{N}\sum_{\vec{k}}
\frac{U^2 \Delta\xi_{\vec{k}}}{(\xi_{\vec{k}}^2+4U^2 \Delta^2)^{3/2}}\,,
\label{eq:deltadelta} \\
\delta n_d&=&\left(\delta \Delta -\frac{V_0}{U}\right)\frac{1}{N}
\sum_{\vec{k}}\frac{2U^2 \Delta\xi_{\vec{k}}}{(\xi_{\vec{k}}^2+4U^2 \Delta^2)^{3/2}}+
\nonumber\\
&&+4\delta n_d\frac{1}{N}\sum_{\vec{k}}
\frac{U^3 \Delta^2}{(\xi_{\vec{k}}^2+4U^2 \Delta^2)^{3/2}}\,.
\label{eq:deltand}
\end{eqnarray}
Since in the limit $\vec{q} \rightarrow 0$ (we write $0$ instead 
of $\{0,0 (,0)\}$ for brevity) Eq. (\ref{eq:Yn}) yields
\begin{eqnarray}
Y_0&=&\frac{1}{2N}\sum_{\vec{k}}
\frac{U^2\Delta\xi_{\vec{k}}}{(\xi_{\vec{k}}^2+4U^2 \Delta^2)^{3/2}}
\,,\\
Y_1&=&-\frac{2}{N}\sum_{\vec{k}}
\frac{U^3\Delta^2}{(\xi_{\vec{k}}^2+4U^2 \Delta^2)^{3/2}}\,, 
\end{eqnarray}
Eqs. (\ref{eq:deltadelta}-\ref{eq:deltand}) can be re-written as
\begin{eqnarray}
4 Y_0(0)\cdot\delta \Delta- [1+2Y_1(0)]\cdot\delta n_d&=& 
4 Y_0(0)\cdot\frac{V_0}{U}\,, \nonumber \\
2 Y_1(0)\cdot\delta \Delta+ 4Y_0(0)\cdot\delta n_d &=& 
[1+2 Y_1(0)]\cdot\frac{V_0}{U}\,. \nonumber 
\end{eqnarray}
This yields
\begin{equation}
\frac{\partial \Delta}{\partial V_0} \equiv 
\frac{\delta \Delta}{V_0} = 
\frac{16 [Y_0(0)]^2+ [1+ 2Y_1(0)]^2}{\{16 [Y_0(0)]^2+4 [Y_1(0)]^2
+2Y_1(0)\}U}\,.
\end{equation} 
This derivative must be negative, as a thermodynamic stability condition [with
the quantity $-V_0$ playing the role of an external field, 
cf. Eq. (\ref{eq:pert})].
On the other hand, it has the same sign as the quantity $D_0(\vec{q})$
at $ \vec{q}\rightarrow 0$, which (owing to $Z_{1,2} \rightarrow 0$ 
in this limit) is given by
\[
D_0( \vec{q}\rightarrow 0)=16 [Y_0(0)]^2+4 [Y_1(0)]^2
+2Y_1(0)\,.
\]

\section{Expressions for $D_1$} 
\label{app:D1}

To leading order
in perturbation, the quantity $D_1(\vec{q})$, introduced in
the main text, has the form
\begin{equation} 
D_1(\vec{q}) = t^\prime T(\vec{q})+ \left\{ \begin{array}{ll}
 {\displaystyle \frac{V_0+V_1E_{rd}}{U \Delta} 
H_0(\vec{q})+ V_1 H_1(\vec{q}),}  &\mbox{even,} \\ ~&~\\ V_2^2 H_2(\vec{q})\,,
&\mbox{odd,} \end{array} \right.
\end{equation}
(again depending on the relative parity of the two bands), where
\begin{widetext}  
\begin{eqnarray}
T (\vec{q})=&& \frac{1}{\Delta}(1-2n_d)\left[\alpha(Y_1^2+4Y_0^2)-2 Y_1 Z_1
-4Y_0\right]+2(Y_1^2+4Y_0^2)(8Y_1-8\alpha Y_0+2\sigma_1-Z_4)+\nonumber \\
&&+ 4(Y_1Z_1+2Y_0)Z_3
-2Y_1Z_2^2+2(1+10 Y_1^2+32Y_0^2-2\alpha Y_0 Y_1+4Y_1)Z_2+2(2-\alpha Y_0+14Y_1)Z_1^2 +\nonumber \\ 
&&+2\alpha(12Y_0^2- Y_1^2)Z_1 
-32(Y_1-1)Y_0Z_1+2(\alpha Y_0 
-4\alpha Y_1 Y_0+4Y_1^2)+\bigg\{(1+2Y_1-Y_1Z_2)\sigma_1+
 \nonumber \\
&&\left.+4(Y_1^2+4Y_0^2)[\sigma_3-2\alpha\sigma_2+(3+\alpha^2)\sigma_1]
+4[\alpha+\sigma_2-\alpha\sigma_1-\frac{1}{\Delta}(1-2n_d)](2Y_0+Y_1 Z_1) \right\} 
\frac{\epsilon_{\vec{q}}+d}{d}\,, \label{eq:Tq}\\ 
H_0(\vec{q})=&&4\left(\frac{\alpha}{2\Delta}(2n_d-1)-\sigma_1-4 Y_1 \right)  (4Y_0^2+Y_1^2) 
-(24Y_0^2
+10 Y_1^2+Y_1)Z_2 - \nonumber \\
&&-(16Y_0-32Y_0Y_1
+10Z_1Y_1)Z_1+1+32 Y_0^2 +4Y_1  
+2(1-2n_d)\frac{Y_1 Z_1+2Y_0}{\Delta}, \\
H_1(\vec{q})=&& \frac{1}{\Delta}(1-2n_d)\left[-Y_1 Z_2-Z_1^2-
4\alpha Y_1 Z_1+ (3\alpha^2+8)(Y_1^2+4Y_0^2)+2Y_1-8\alpha Y_0\right]
-2(Y_1^2+4Y_0^2) \times \nonumber \\
&&\times(Z_3+32Y_0)-16(Y_1^2-4Y_0^2)Z_1-8Y_0(1+2Y_1)Z_2+2Z_1^3-16Y_0Z_1^2
-2(1+8Y_1)Z_1-\nonumber \\
&&-16Y_0(1+2Y_1)
- 8(Y_1Z_1+2Y_0)\sigma_1+6(Y_1^2+4Y_0^2)(2\alpha\sigma_1-\sigma_2)-(Y_1^2+4Y_0^2)\frac{\epsilon_{\vec{q}}-2d}{2U^2\Delta^3}\,.
\label{eq:H1q}
\end{eqnarray}
Here,
\begin{equation}
\alpha=\frac{E_{rd}}{U\Delta}\,,\,\,\,\sigma_n=\frac{1}{U^n \Delta^{n+1} N}
\sum_{\vec{k}}\left(\epsilon _{\vec{k}} \right)^n n^d_{\vec{k}}\,,
\end{equation}
and the quantities
\begin{eqnarray}
Z_3(\vec{q})=&&\frac{2-4n_d}{\Delta} +\frac{3 \alpha^2}{2 \Delta}
-\frac{\epsilon_{\vec{q}}-2d}{4U^2\Delta^3}+Y_4(\vec{q})-16Y_0(\vec{q})
-3\alpha^2 \frac{n_d}{\Delta}+6\alpha\sigma_1-3\sigma_2
\,,\\
Z_4(\vec{q})=&&2\frac{\alpha^3}{\Delta}-\alpha
\frac{\epsilon_{\vec{q}}-2d}{U^2\Delta^3}+\frac{2\alpha}{\Delta}-8
+Y_5(\vec{q})-16Y_1(\vec{q})-4\alpha(1+\alpha^2)\frac{n_d}{\Delta}
+4(3\alpha^2+1)\sigma_1-12\alpha\sigma_2+4\sigma_3\,,
\end{eqnarray}
vanish in the long-wavelength limit, $\vec{q} \rightarrow0$, for the 
unperturbed case. On the r.\ h.\ s. of Eqs. (\ref{eq:Tq}--\ref{eq:H1q}),
we omitted the argument $\vec{q}$ of the functions $Y_n(\vec{q})$ and
$Z_n(\vec{q})$; note that these are computed using the leading-order
expressions $\Delta_{\vec{k}}$ and $n^d_{\vec{k}}$, Eqs. 
(\ref{eq:Delta}--\ref{eq:nd}).

Since the general expression for $H_2(\vec{q})$ is rather lengthy, here
we restrict ourselves to the case of $\vec{q}=\vec{Q}_0$:
\begin{eqnarray} 
H_2(\vec{Q}_0)=&&-\frac{1}{2}\left[ (4+\alpha^2)I_1+I_2 \right]\left\{\left[1-2\alpha
Y_0(\vec{Q}_0)\right]^2+16 \left[Y_0(\vec{Q}_0)\right]^2\right\}\,, \\
I_1=&&\frac{1}{U\Delta^2N}\sum_{\vec{k}}\frac{\lambda_{\vec{k}}^2 n^d_{\vec{k}}
}{\epsilon_{\vec{k}}}\,\,,\,\,\,\,\,\,I_2=\frac{1}{U^3\Delta^4 N}
\sum_{\vec{k}}\epsilon_{\vec{k}}\lambda_{\vec{k}}^2 n^d_{\vec{k}}\,.
\end{eqnarray}
\end{widetext}
It is easy to see\cite{sign} that $I_{1,2}>0$, 
hence $H_2(\vec{Q}_0)$ is negative.

\section {Further details on Figs. 1 and 2} 
\label{app:figdetail}

\begin{figure}
\includegraphics{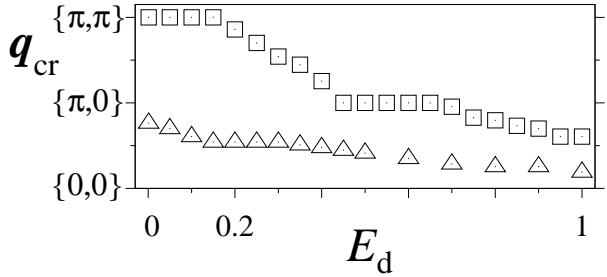}
\caption{\label{fig:qcr} Values of $\vec{q}_{cr}$ in the $t^\prime \neq 0$ case
for a 2D EFKM with $U=1$ (squares) and $U=0.5$ (triangles). 
The corresponding values of $t^\prime_{cr}$ are plotted in Fig. 2 
(solid lines in the two right panels).}
\end{figure}

As the strength of the
perturbation, Eq. (4), is decreased towards the corresponding critical value,
$t^\prime_{cr}$ or $V_{i,cr}$, the collective excitation spectrum 
softens at a certain $\vec{q}=\vec{q}_{cr}$, {\it viz.},
$\omega_{\vec{q}_{cr}}=0$, signalling an instability of the excitonic 
insulator. 
The spectra shown in Fig. 1 correspond to 
$\vec{q}_{cr}=\vec{Q}_0\equiv \{\pi,\pi\}$. While this is indeed a 
typical case, other
situations are possible, especially at smaller $U$. This is illustrated
in Fig. \ref{fig:qcr}. In  the momentum space, the line $\{0,0\}-\{\pi,0\}
-\{\pi,\pi\}-\{0,0\}$ was probed, and $\vec{q}_{cr}$ was found to lie
on the  $\{0,0\}-\{\pi,0\}-\{\pi,\pi\}$ segments.

\begin{thebibliography}{999}
\bibitem{sign} Note that the quantity $n^d_{\vec{k}}$, Eq. (\ref{eq:nd}), 
depends on momentum
via $\epsilon_{\vec{k}}$, and increases as the latter varies from $-d$ to $d$.
\bibitem{recover} When all $V_i$ vanish, we recover, in the appropriate limit,
Eq. (4.17) of Ref. \onlinecite{Jerome}.
\bibitem{crossover} This is replaced with a steep crossover when either
$V_0$ or $V_1$ differs from zero, similar to  ferro- to paramagnetic 
transition in a weak applied field.
\bibitem{firstorder} A first-order 
transition at $T_1< T_c$ may also occur, especially when lattice 
effects\cite{Kikoin83}
are taken into account.
\end{thebibliography}
\end{document}